\documentclass[sigconf]{acmart}
\usepackage{tikz}
\usetikzlibrary{shapes, arrows, positioning}

\AtBeginDocument{%
  }

\setcopyright{cc}
\copyrightyear{2026}
\acmYear{2026}
\acmDOI{}
\acmConference[CHI 2026 Workshop on Next Steps for Augmented Reality On-the-Move]{CHI 2026 Workshop on Next Steps for Augmented Reality On-the-Move: Challenges \& Opportunities}{Aril 2026}{Barcelona, Spain}
\acmISBN{}

\begin{document}

\title{Designing Augmented Reality for Preschoolers on the Move}

\author{Supriya Khadka}
\affiliation{%
  \institution{George Mason University}
  \city{Fairfax, Virginia}
  \country{USA}}
\email{skhadk@gmu.edu}

\author{Sanchari Das}
\affiliation{%
  \institution{George Mason University}
  \city{Fairfax, Virginia}
  \country{USA}}
\email{sdas35@gmu.edu}

\renewcommand{\shortauthors}{Khadka and Das}

\begin{abstract}
Advancements in augmented reality (AR) technologies offer immense potential for mobile experiences. However, most commercial and educational AR systems assume a baseline of predictable user behavior and stationary interaction. Preschoolers and children in early childhood education, specifically ages 3 to 8, are naturally erratic, physically dynamic, and prone to rapid locomotion, making them the ultimate stress test for mobile spatial computing. Through a focused analysis of recent literature on physical activity and spatial learning in AR for preschoolers, this paper identifies points of friction in current mobile deployments. We highlight recurring failures in camera tracking during dynamic movement, physical safety hazards caused by screen-induced distraction, spatial crowding around physical markers, and the privacy risks of continuous environmental surveillance.  To address these challenges, we propose AnchorPlay AR, a conceptual prototype for a privacy-preserving, audio-first spatial application. By explicitly separating locomotion from visual tracking, AnchorPlay AR uses audio cues to safely guide movement and reserves visual augmentation for stationary moments, offering a safer framework for preschoolers in constant motion.
\end{abstract}


  

\maketitle

\section{Introduction}
The vision of ubiquitous augmented reality (AR) requires systems supporting interaction in highly dynamic settings. Preschoolers, specifically children ages 3 to 8, are naturally prone to rapid, unpredictable locomotion~\cite{cools2009movement}. Hence, AR has been explored extensively in early childhood education as a tool to leverage their natural curiosity and physical energy~\cite{khadka2026sok}. However, young children lack fully developed ergonomic control~\cite{yilmazExaminationVocabularyLearning2022}. This creates an embodied mismatch between the child's physical capabilities and the system's demands~\cite{khadka2026sok, khadka2026xr}. While recent efforts utilize AR for physically active spatial learning~\cite{stearneAugmentedRealityPlaygrounds2025, kurniawanARtraceAugmentedReality2019, liangHandGesturebasedInteractive2017}, they repeatedly encounter technical failures when children move, as traditional AR forces stationary, precision-focused paradigms onto hyper-mobile users.

Beyond physical ergonomics, deploying mobile AR for this demographic introduces privacy vulnerabilities~\cite{khadka2026sok, noah2021exploring, duezguen2020towards, noah2025pins, reddington2022development}. Systems that require an always-on camera feed to navigate physical spaces continuously capture sensitive environmental data and bystander faces. For a vulnerable age group governed by strict privacy standards, minimizing continuous visual data collection is a fundamental security imperative~\cite{noah2022security, kishnani2024dual, adhikari2025natural}. Therefore, the field urgently needs a reliable design approach for secure, active learning. Building on our prior work evaluating AR deployments in real-world early childhood settings~\cite{khadka2026sok}, we review recent literature pinpointing these exact physical, technical, and security barriers. We then propose AnchorPlay AR, a conceptual interaction model designed specifically to resolve those issues by separating auditory locomotion from visual tracking.

\begin{figure*}[ht]
\centering
\includegraphics[width=0.72\linewidth]{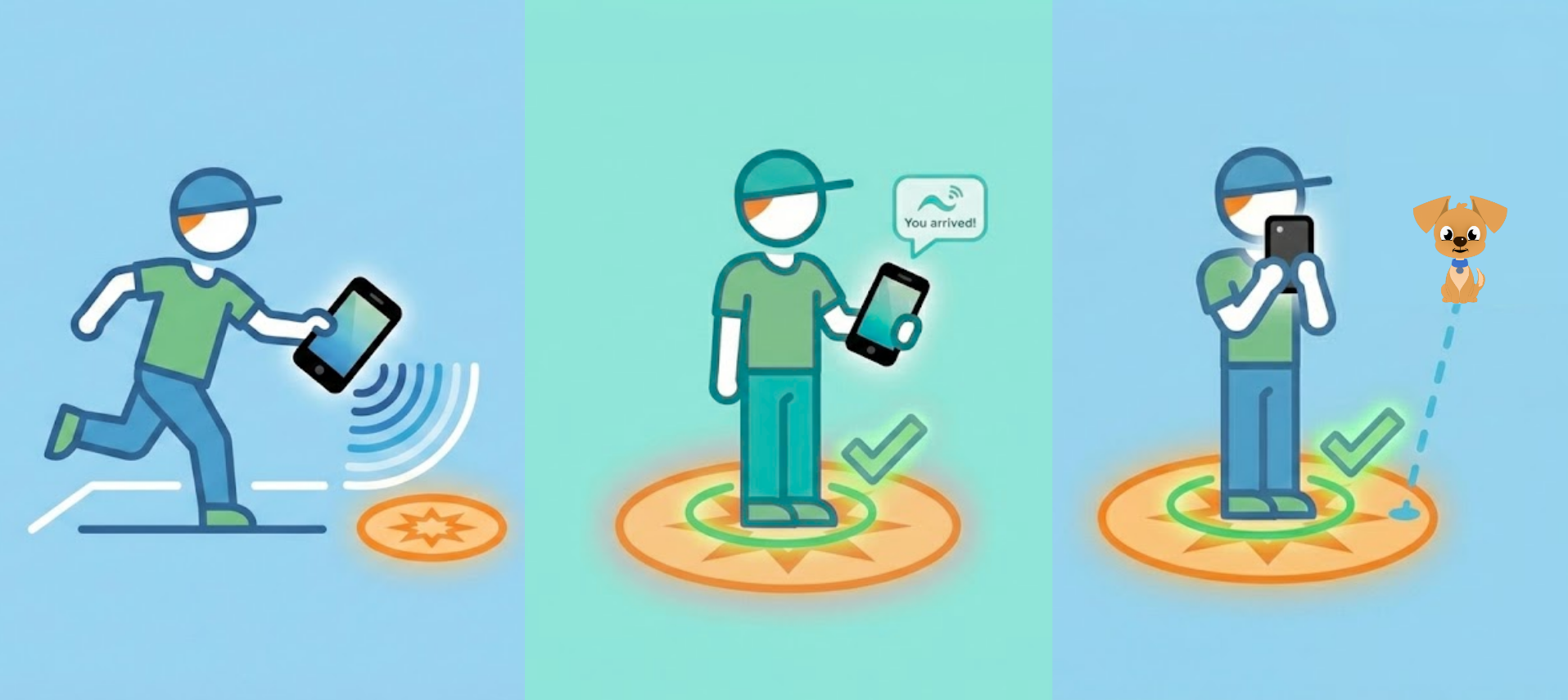} 
\caption{A three-panel storyboard illustrating the stop-and-look interaction model of AnchorPlay AR. Audio cues guide the preschooler safely across the physical space (Phase 1). Visual AR tracking, represented by the puppy, only activates when the child stands completely still on a physical anchor (Phases 2 and 3). This design explicitly separates locomotion from screen viewing, reducing tracking loss and physical collision risks.}
\label{fig:stop-and-look-model}
\Description{A three panel storyboard showing a preschooler using the AnchorPlay AR system. Panel 1 shows a child walking with a tablet held down by their side. Sound waves emerge from the device, and a target circle is on the floor ahead. Audio cues guide the child safely. Camera tracking is paused. Panel 2 shows the child standing completely still on the target circle. The tablet is still held down, and a green checkmark appears at their feet to indicate arrival. The text below says Physical anchor detected. Locomotion registers as stopped. Audio prompts the user to look. Panel 3 shows the child standing on the target while holding the tablet up in front of their face. A dashed line connects the tablet to a virtual 3D puppy floating directly above the target circle. Device tracking activates only when stable, reducing visual fatigue and tracking loss.}
\end{figure*}

\section{The Reality of AR-on-the-move for Preschoolers}
We begin our analysis by drawing from a comprehensive review of 85 studies focusing on AR in early childhood education~\cite{khadka2026sok}. A recurring theme across this broader dataset is that AR interventions frequently lose their effectiveness due to the natural, erratic movement and developing ergonomic control of children ages 3 to 8. To deeply investigate this mobility problem, we narrow our focus to a specific subset of 9 studies that explicitly target AR for physical activity and spatial learning. Through this targeted analysis, we identify four primary failure points when AR is deployed in motion for preschoolers.

\textbf{Physical Safety and Ergonomic Fatigue.}
Mobile AR often forces users to look through a screen while navigating physical environments. A laboratory study noted children ages 5 to 8 risk tripping or colliding with objects while moving with smartphone AR~\cite{stearneAugmentedRealityPlaygrounds2025}. Furthermore, the ergonomic demand of maintaining a specific camera angle during locomotion is high~\cite{kulasekaraGameCentricElearning2025}. The co-design of the \textit{AR Adventures} spatial learning app revealed that holding a tablet upright creates arm fatigue for preschoolers~\cite{lewis-presserDesigningAugmentedReality2025}. Studies also found interactions requiring physical movement and perspective shifts induced physical strain in young children, making it difficult to interact with the learning material~\cite{raduComparingChildrensCrosshair2016, lewis-presserDesigningAugmentedReality2025}.

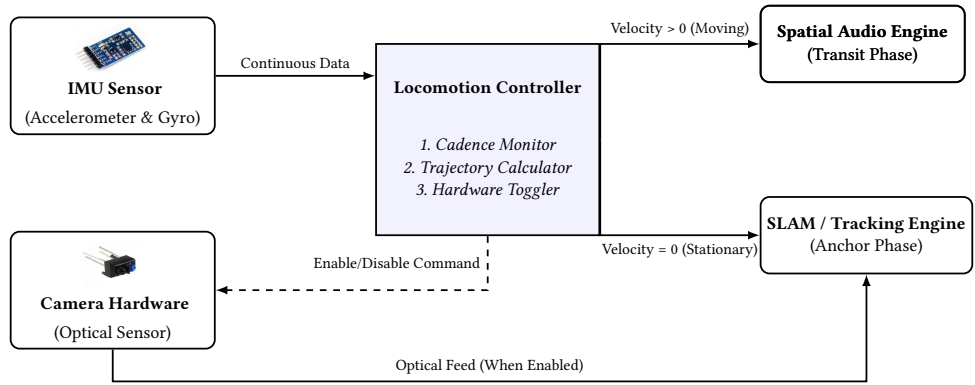
\begin{figure*}[ht]
\centering
\resizebox{0.72\linewidth}{!}{%
\begin{tikzpicture}[
    node distance=1.5cm and 2.5cm,
    block/.style={draw, rectangle, rounded corners, align=center, minimum width=3.2cm, minimum height=1.2cm, font=\small, thick},
    controller/.style={draw, rectangle, align=center, minimum width=3.5cm, minimum height=3cm, fill=blue!5, thick, font=\small},
    arrow/.style={-latex, thick}
]

\node[block] (imu) {
    \includegraphics[width=1.2cm]{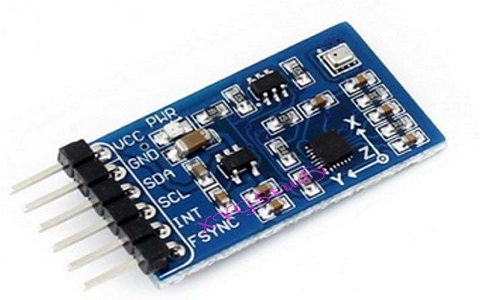} \\ 
    \vspace{0.1cm} 
    \textbf{IMU Sensor}\\(Accelerometer \& Gyro)
};

\node[block, below=of imu] (cam) {
    \includegraphics[width=1.2cm]{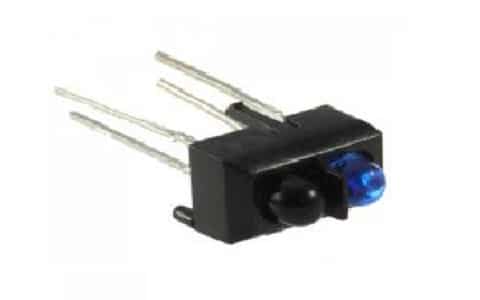} \\ 
    \vspace{0.1cm} 
    \textbf{Camera Hardware}\\(Optical Sensor)
};

\node[controller, right=of imu, yshift=-1cm] (loc) {\textbf{Locomotion Controller}\\ \vspace{0.2cm} \\ \textit{1. Cadence Monitor}\\ \textit{2. Trajectory Calculator}\\ \textit{3. Hardware Toggler}};

\node[block, right=of loc, yshift=1.5cm] (audio) {\textbf{Spatial Audio Engine}\\(Transit Phase)};
\node[block, right=of loc, yshift=-1.5cm] (slam) {\textbf{SLAM / Tracking Engine}\\(Anchor Phase)};

\node[block, right=of loc, yshift=1.5cm] (audio) {\textbf{Spatial Audio Engine}\\(Transit Phase)};

\draw[arrow] (imu.east) -- node[above, font=\footnotesize] {Continuous Data} (imu.east -| loc.west);

\draw[arrow, dashed] (loc.south) |- node[near start, left, font=\footnotesize] {Enable/Disable Command} (cam.east);

\draw[arrow] (cam.south) -- ++(0,-0.5) -| node[near start, above, font=\footnotesize] {Optical Feed (When Enabled)} (slam.south);

\draw[arrow] (loc.east) |- node[near end, above, font=\footnotesize] {Velocity > 0 (Moving)} (audio.west);
\draw[arrow] (loc.east) |- node[near end, below, font=\footnotesize] {Velocity = 0 (Stationary)} (slam.west);

\end{tikzpicture}
}
\caption{System architecture of AnchorPlay AR. The central Locomotion Controller continuously processes IMU telemetry. During transit, it routes trajectory data exclusively to the Spatial Audio Engine. The system only issues an enable command to the Camera Hardware and SLAM engine when a stationary state (Velocity = 0) is achieved.}
\label{fig:tech_arch}
\end{figure*}

\textbf{Tracking Loss During Dynamic Locomotion.} Current mobile AR struggles to maintain fidelity during sudden, erratic movement. Research comparing crosshair and finger interactions in handheld AR found that younger children, specifically ages 5 to 6, frequently lost tracking because they could not keep the camera steadily pointed at targets while walking~\cite{raduComparingChildrensCrosshair2016}. Similarly, the \textit{Kidstac} e-learning application, which utilized pose detection to encourage physical activity, reported a drop in system accuracy when preschoolers performed dynamic poses like jumping compared to stable poses~\cite{kulasekaraGameCentricElearning2025}. Hardware tracking limits also pose a challenge. For example, Leap Motion tracking in interactive puppetry systems frequently failed when children naturally moved their hands out of the restricted tracking zone during energetic play~\cite{liangHandGesturebasedInteractive2017}.

\textbf{Spatial Chaos and Crowding.} Deploying AR in shared physical spaces introduces social and environmental complexities. An investigation into student engagement found that placing physical AR markers around a room led to high crowding, frustration, and physical pushing as multiple students attempted to navigate to the same spatial anchor simultaneously~\cite{drljevicInvestigatingDifferentFacets2022}. When children are expected to move through a room to find digital content, the physical clustering creates a chaotic environment that detracts from the intended spatial learning outcomes~\cite{gecu-parmaksizEffectAugmentedReality2020a, raduComparingChildrensCrosshair2016}.

\textbf{Privacy and Security Risks of Continuous Tracking}
Beyond physical and technical limitations, designing AR for locomotion introduces multiple privacy vulnerabilities. Traditional AR navigation relies on an always-on camera feed to map the environment and track the user's position~\cite{khadka2026sok}. In early childhood education settings, this continuous visual data collection poses a severe security risk~\cite{lewis-presserDesigningAugmentedReality2025}.  The camera inadvertently captures sensitive environmental data, classroom layouts, and the faces of bystander children who may not be part of the learning activity. Minimizing this continuous surveillance is a fundamental requirement for deploying safe, ethical spatial computing for this demographic.

\section{AnchorPlay AR}
We propose AnchorPlay AR, a conceptual prototype managing AR locomotion for preschoolers through a privacy-preserving, audio-first interaction model. It acts as a distributed scavenger hunt where visual AR triggers only when the user achieves a safe, stationary posture. This directly resolves the compounded challenges of physical fatigue, tracking loss, spatial crowding, embodied mismatch, and privacy vulnerabilities.

The system relies on a \textbf{stop-and-look interaction model} (Figure~\ref{fig:stop-and-look-model}). Children receive instructions guiding their locomotion via a spatial audio framework, which adjusts cues based on IMU-calculated trajectories. During this transit phase, low-power motion sensors act as hardware gatekeepers. The system exclusively polls the accelerometer and gyroscope, programmatically suspending the spatial tracking engine. Holding the device down with the camera disabled explicitly separates locomotion from screen viewing. This mitigates tripping hazards, eliminates continuous computational overhead, prevents tracking drift from swinging devices, and enforces the privacy imperative by shutting off the camera feed during classroom navigation.

Transitioning to visual AR relies on \textbf{physical proximity as the primary interaction mechanism}, effectively resolving the embodied mismatch of traditional AR. The device IMU continuously monitors step cadence (Figure~\ref{fig:tech_arch}). Once it detects a complete cessation of gross motor movement for a predefined threshold, the software initializes the camera and visual tracking engine. The system then performs a localized search for a physical floor anchor. Because the child and device are completely stationary, the spatial map initializes instantly without motion blur. Once the marker is recognized, the digital reward is instantiated. If locomotion resumes, the state machine immediately terminates visual tracking, ensuring movement and camera tracking remain mutually exclusive states. Finally, the application utilizes \textbf{distributed waypoints} to dynamically assign different paths, ensuring users remain physically scattered to prevent chaotic crowding.

\section{Reflections and Opportunities}
Designing AR on the move for preschoolers exposes the severe limitations of interfaces expecting stationary precision. Moving forward, we must rethink how we distribute interaction across multiple sensory modalities. Relying solely on visual AR while navigating a physical space dramatically increases the risks of collision, ergonomic fatigue, and technical tracking failure. Embracing multimodal paradigms, like AnchorPlay AR, allows spatial audio to guide locomotion while visual AR serves exclusively as a destination reward. This approach actively protects the child while heavily reducing continuous computational load during erratic movement. Furthermore, designing for this unpredictable demographic provides a rigorous stress test for the broader pursuit of ubiquitous computing. If a spatial computing architecture can safely guide a running preschooler without tracking loss or physical injury, it will inherently yield highly reliable experiences for adult users navigating busy urban environments. Ultimately, meaningful augmentation requires systems that adapt seamlessly to human movement, rather than forcing the human body to conform to tracking constraints.

\begin{acks}
    We acknowledge the Data Agency and Security (DAS) Lab at George Mason University, where this study was conducted. This research was supported in part by an unrestricted gift from Google. The opinions expressed in this work are solely those of the authors.
\end{acks}

\bibliographystyle{ACM-Reference-Format}
\bibliography{main}

\end{document}